\documentclass[aps,prb,twocolumn,showpacs,superscriptaddress,nofootinbib]{revtex4-1}  % for review and submission
\usepackage{graphicx}  % needed for figures
\usepackage{dcolumn}   % needed for some tables
\usepackage{bm}        % for math
\usepackage{amsmath, amsthm, amssymb}
\usepackage{bbold} % to generate identity 1
\usepackage{mathrsfs}
\usepackage{array}
\usepackage[usenames]{color}
\usepackage{soul} % to use \st for strikeout 
\usepackage{ulem} % to use \sout for strikeout to include equations
\begin{document}

\normalem

\title{Ferromagnetism beyond Lieb's theorem}
\author{Natanael C. Costa}
\affiliation{Instituto de Fisica, Universidade Federal do Rio de
Janeiro Cx.P. 68.528, 21941-972 Rio de Janeiro RJ, Brazil}
\author{Tiago Mendes-Santos}
\affiliation{Instituto de Fisica, Universidade Federal do Rio de
Janeiro Cx.P. 68.528, 21941-972 Rio de Janeiro RJ, Brazil}
\author{Thereza Paiva}
\affiliation{Instituto de Fisica, Universidade Federal do Rio de
Janeiro Cx.P. 68.528, 21941-972 Rio de Janeiro RJ, Brazil}
\author{Raimundo R. dos Santos} 
\affiliation{Instituto de Fisica, Universidade Federal do Rio de
Janeiro Cx.P. 68.528, 21941-972 Rio de Janeiro RJ, Brazil}
\author{Richard T. Scalettar}
\affiliation{Department of Physics, University of California, Davis, CA 95616,
USA}
\begin{abstract}
%Colour code: {\color{blue} Blue = text needs attention.}
%{\color{red} Red = Notes, queries, etc. for ourselves.}
The noninteracting electronic structures of tight binding models on
bipartite lattices with unequal numbers of sites in the two sublattices
have a number of unique features, including the presence of spatially
localized eigenstates and flat bands.  When a \emph{uniform} on-site
Hubbard interaction $U$ is turned on, Lieb proved rigorously that at
half filling ($\rho=1$) the ground state has a non-zero spin.
In this paper we consider a `CuO$_2$ lattice (also known as `Lieb
lattice', or as a decorated square lattice), in which `$d$-orbitals'
occupy the vertices of the squares, while `$p$-orbitals' lie halfway
between two $d$-orbitals; both $d$ and $p$ orbitals can accommodate only
up to two electrons.  
We use exact Determinant Quantum Monte Carlo (DQMC) simulations to
quantify the nature of magnetic order through the behavior of
correlation functions and sublattice magnetizations in the different
orbitals as a function of $U$ and temperature; we have also calculated
the projected density of states, and the compressibility.  We study both
the homogeneous (H) case, $U_d= U_p$, originally considered by Lieb, and
the inhomogeneous (IH) case, $U_d\neq  U_p$.  
For the H case at half filling, we found that the global magnetization rises sharply at weak coupling, and then stabilizes towards the strong-coupling (Heisenberg) value, as a result of the interplay between the ferromagnetism of like sites and the antiferromagnetism between unlike sites; we verified that the system is an insulator for all $U$.  
For the IH system at half filling, we argue that the case $U_p\neq U_d$ falls under Lieb's theorem, provided they are positive definite, so we used DQMC to probe the cases $U_p=0,U_d=U$ and $U_p=U, U_d=0$.  
We found that the different
environments of $d$ and $p$ sites lead to a ferromagnetic insulator when
$U_d=0$; by contrast, $U_p=0$ leads to to a metal without any magnetic
ordering.  In addition, we have also established that at density
$\rho=1/3$, strong antiferromagnetic correlations set in, caused by the
presence of one fermion on each $d$ site. 
\end{abstract}

\date{Version 2.15 -- \today}

\maketitle

%%%%%%%%%%%%%%%%%%%%%%%%%%%%%%%%%%%%%%%%%%%%%%%%%%%%%%%%%%%%%%%%%%
%%%%%%%%%%%%%%%%%%%%%%%%%%%%%%%%%%%%%%%%%%%%%%%%%%%%%%%%%%%%%%%%%%
\section{Introduction}
%%%%%%%%%%%%%%%%%%%%%%%%%%%%%%%%%%%%%%%%%%%%%%%%%%%%%%%%%%%%%%%%%%
%%%%%%%%%%%%%%%%%%%%%%%%%%%%%%%%%%%%%%%%%%%%%%%%%%%%%%%%%%%%%%%%%%

Within early mean field theories (MFT's), the ground state of the single band Hubbard Hamiltonian,\cite{hubbard63} e.g.~on a square lattice, was predicted to support both long range ferromagnetism (FM) and anti-ferromagnetism (AFM), with the two ordering wave vectors each occupying broad regions in the density ($\rho$)--interaction strength ($U$) phase space.\cite{hirsch85,fazekas99}  
However, when treated with more accurate methods like Quantum Monte Carlo (QMC) simulations and generalized Hartree-Fock approaches,\cite{Bach1994,Bach1997} this parity is broken. 
FM proves to be much more elusive,\cite{hirsch89,white89} and indeed seems to be entirely absent from the square lattice phase
diagram\cite{rudin85} except in `extreme' situations such as the Nagaoka
regime of doping with a single electron away from half-filling at very
large $U$ (many times the kinetic energy bandwidth).\cite{hanisch93,[{In
continuum models it also has been argued that the ratio of interaction
strength to kinetic energy needs to be an order of magnitude larger than
that suggested by MFT; see, e.g.,\,}] zong02}  The difficulty in
achieving FM in the Hubbard Hamiltonian is unfortunate, since its
explanation was one of the original motivations of the
model.\cite{hubbard63,[{Within a MFT treatment, the condition for FM in
the Hubbard model is equivalent to that due to Stoner,\,}] stoner38}

How, then, might itinerant ferromagnetism be achieved in a model
Hamiltonian?  One route retains a single band but introduces frustration
({\it e.g.}~through next near neighbor hopping) which shifts spectral
weight to the band edges and minimizes the kinetic energy cost of the
magnetic state.\cite{mielke91,ulmke98,vollhardt01} Additional interaction terms
such as next-neighbor direct exchange
\cite{hubbard63,hirsch89b,strack94,wahle98} or bond-charge (correlated
hopping) can also increase ferromagnetic tendencies, at least within
MFT\cite{amadon96} or Gutzwiller approximation\cite{kollar01}
treatments.

A second route to ferromagnetism is through the presence of several
electronic bands.  Within one picture, the resulting Hund's rule
interactions play a crucial role.\cite{held98,Mielke1993}  A distinct scenario, and
the one we carefully explore here, focusses instead on the presence of
special noninteracting dispersion relations.  In this context, a series
of rigorous results were obtained.  First, Lieb\cite{lieb89,*lieb89err} established
a theorem stating that in a class of bipartite geometries in any spatial
dimension, with unequal numbers of sites, $N_\mathcal{A}$ and
$N_\mathcal{B}$, in the two sublattices ($\mathcal{A}$ and
$\mathcal{B}$), the ground state has total spin
$S=|N_\mathcal{A}-N_\mathcal{B}|/2$.  The class of bipartite lattices
for which the theorem was originally proved was subject to the following
restrictions: the Hubbard repulsion $U$ must be the same on every
lattice site; hopping $t_{ij} c_{i}^{\dagger}c_{j}^{\phantom{\dagger}}$
can only take place between sites $ij$ in opposite sublattices; there
can be no single-particle chemical potential terms
$\varepsilon^{\phantom{\dagger}}_{i}
c_{i}^{\dagger}c_{i}^{\phantom{\dagger}}$.  With these conditions, the
Hamiltonian $\mathcal{H}$ is particle-hole symmetric (PHS), and each
site, irrespective of being on the ${\cal A}$ or the ${\cal B}$
sublattice, is exactly half-filled.  
One should note, however, that Lieb himself warned that ``spatial ordering is not implied'' by a
non-vanishing total spin; in addition, here the use of \emph{ferromagnetism} should be understood as encompassing 
\emph{unsaturated ferromagnetism}, though some authors (see the Erratum to Ref.\,\onlinecite{lieb89}) 
advocate the use of \emph{ferrimagnetism} in this case.
A subsequent development was
achieved\cite{shen94} by establishing that spin-spin correlation
functions $\langle \Phi_0|\mathbf{S}_i\cdot \mathbf{S}_j|\Phi_0\rangle$,
where $|\Phi_0\rangle$ is the ground state, are positive (negative) for
$i$ and $j$ on the same (different) sublattices; here, again, long-range
order is not necessarily implied.\cite{tasaki98,tasaki98b}

Possible ferromagnetic order is closely tied to the fact that, in the non-interacting limit, tight-binding Hamiltonians on lattices with this geometry and obeying these conditions, have highly degenerate localized eigenstates, from which linear combinations can be constructed to form a perfectly flat electronic band.  
At half-filling for the entire lattice, this flat band itself is
precisely half-filled.  Lieb's theorem was subsequently generalized to
other graphs including the Kagom\'e and the square lattice with
cross-hoppings on half the squares.\cite{mielke91}  It is notable that
one of the essential ingredients in this route to ferromagnetism, PHS,
is precisely what is broken in other scenarios such as the introduction
of frustration.  
The implications of the Lieb theorem have also been explored for more general geometries,\cite{mielke91,tasaki92}  and one should note that flat band ferromagnetism is independent of lattice dimensionality.\cite{Mielke1999a,Mielke1999b}
FM was found to occur away from the singular flat-band limit, i.e.\,in models
with the perfect cosine dispersion characteristic of the Hubbard model
with near neighbor hopping only on linear, square, and cubic
lattices.\cite{tasaki95}

\begin{figure}[t]
\includegraphics[scale=0.7]{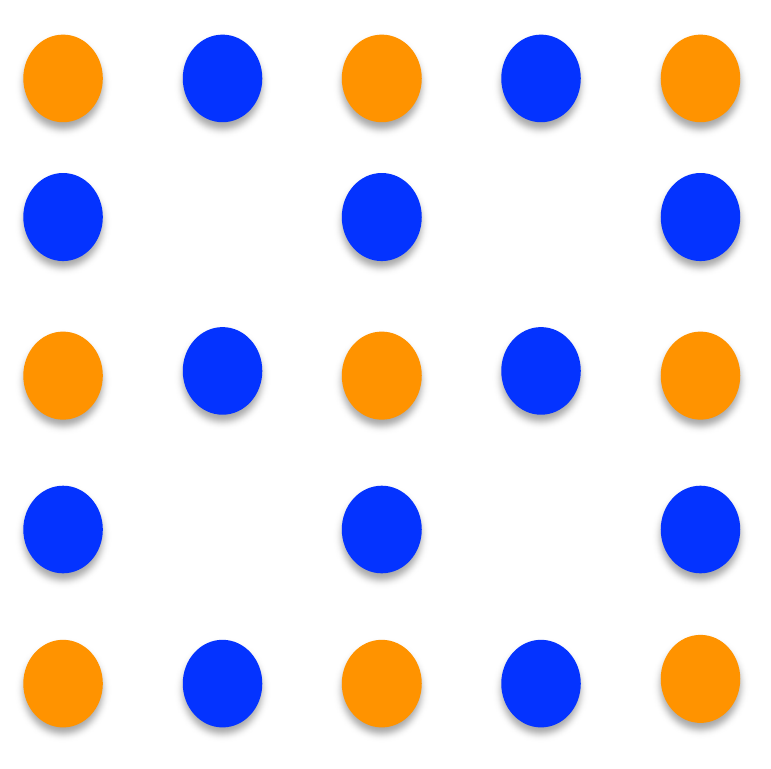} 
\caption{(Color online) 
The Lieb lattice (or CuO$_2$ lattice). The four-fold coordinated
$d$-sites appear in lighter color (orange) and belong to the
$\mathcal{A}$ sublattice, while the two-fold coordinated $p$-sites
appear in darker color (blue) and belong to the $\mathcal{B}$
sublattice.  }
\label{fig:CuO2}
\end{figure}

One particular geometry to which Lieb's theorem applies is the `CuO$_2$
lattice', also referred to as the Lieb lattice; see
Fig.\,\ref{fig:CuO2}.  In spite of the similarities with the actual
CuO$_2$ sheets of high-$T_c$ cuprates, one must stress that the relevant
fillings for superconductivity in these materials is one hole per
CuO$_2$ unit cell, rather than half-filling (three holes per unit cell),
and in fact a significant site energy difference $\varepsilon_p -
\varepsilon_d$ exists between occupation of the copper $d$ and oxygen
$p$ orbitals (violating one of the restrictions of Lieb's theorem).\cite{mielke91}
Indeed, the cuprate materials exhibit AFM rather than FM.

Some consequences of the peculiar geometry of the Lieb lattice have been
recently pursued in several theoretical
studies.\cite{Weeks10,Zhao12,Nita13} While these studies did not include
on-site interactions, which are directly linked with ferromagnetism,
some effects of on-site repulsion $U$ have only been investigated with
the aid of Dynamical Mean Field (DMFT):\cite{Noda09,Noda14} 
it was found that each sublattice magnetization behaves monotonically with $U$, and this correlates with the local density of states.  
Finally, experimental
realizations of the Lieb lattice as photonic lattices have been recently
reported,\cite{Vicencio15,Mukherjee15} and one should expect optical
lattices could also be set up with this topology, motivated by the
possibility of engineering ferromagnetic states through the control of
interactions.

In view of this, several issues regarding the existence of
ferromagnetism on the Lieb lattice should be addressed, and  here we use
determinant Quantum Monte Carlo (DQMC) which treats the interacting
electron problem exactly on lattices of finite size.  First, a detailed
analysis of the sublattice-resolved spatial decay of spin correlations
and order parameters would add considerably to the understanding of how
the basic units conspire to yield a robust polarized state.  Secondly,
can ferromagnetism still be found if one deviates from the conditions of
Lieb's theorem, e.g., by relaxing the constraint of uniform $U$, i.e.,
allowing for $U_d\neq U_p$ (on ``oxygen'' and ``copper'' sites,
respectively)?  We then go beyond Lieb's theorem by distinguishing two
situations, namely, the case where both $U_p$ and $U_d$ are non-zero,
and the cases in which the on-site repulsion is switched off on either
$p$ sites or $d$ sites.  Away from half filling, DQMC simulations are
plagued by the infamous `minus-sign problem', which prevents us from
reaching very low temperatures.  Nonetheless, we can still shed some
light into the effects on magnetic ordering by switching off the
repulsion on either $p$ or $d$ sites.

The paper is organized as follows.  In Sec.\,\ref{sec:HQMC} we present
the main features of the Hubbard Hamiltonian on the Lieb lattice, and
highlight the DQMC method together with the quantities of interest.  The
results for the homogeneous and inhomogeneous lattices at half filling
are presented in Secs.\,\ref{sec:homog} and \ref{sec:inhomog},
respectively; the behavior away from half filling is briefly analyzed in
Sec.\,\ref{sec:doped}.  Our main conclusions are then summarized in
Sec.\,\ref{sec:conc}.

%%%%%%%%%%%%%%%%%%%%%%%%%%%%%%%%%%%%%%%%%%%%%%%%%%%%%%%%%%%%%%%%%%
\section{Three Band Hubbard Hamiltonian and Quantum Monte Carlo
Methodology}
\label{sec:HQMC}
%%%%%%%%%%%%%%%%%%%%%%%%%%%%%%%%%%%%%%%%%%%%%%%%%%%%%%%%%%%%%%%%%%

The particle-hole symmetric three band Hubbard Hamiltonian on a Lieb lattice,
\begin{align}
\hat H -\mu \hat N= &-t_{pd} \sum_{{\bf r} \sigma}
\big( \,
d_{{\bf r} \sigma}^{\dagger}   \, p_{{\bf r} \sigma}^{x\phantom{\dagger}}
+ d_{{\bf r} \sigma}^{\dagger}   \, p_{{\bf r} \sigma}^{y\phantom{\dagger}}
+ {\rm h.c.} \big)
%% p_{i \sigma}^{x\dagger}   \, d_{i \sigma}^{\phantom{\dagger}}
%% \, \big)
%% - \mu \sum_{i\sigma} n_{i \sigma}
\nonumber \\
&-t_{pd} \sum_{{\bf r} \sigma}
\big( \,
d_{{\bf r}  \sigma}^{\dagger}   
\, p_{{\bf r} -\hat x \, \sigma}^{x\phantom{\dagger}}
+ d_{{\bf r}  \sigma}^{\dagger}   
\, p_{{\bf r} -\hat y \, \sigma}^{y\phantom{\dagger}}
+ {\rm h.c.} \big)
\nonumber \\
 &+  \sum_{{\bf r} \alpha} U_\alpha
 \left(\, n^{\alpha}_{{\bf r} \uparrow} -\frac12 \, \right)
 \left(\, n^{\alpha}_{{\bf r} \downarrow } -\frac12 \, \right)
\nonumber \\
& +  \sum_{{\bf r} \alpha \sigma} \varepsilon_\alpha n^{\alpha}_{{\bf r} \sigma} 
- \mu  \sum_{{\bf r} \alpha \sigma}  n^{\alpha}_{{\bf r} \sigma} 
\label{eq:ham}
\end{align}
contains inter- and intra-cell hopping $t_{pd}$ between a (`copper')
$d$- and two (`oxygen') $p^x, \, p^y$ orbitals.  In this paper we
consider the on-site repulsion both as homogeneous, $U_p=U_d=U$, in
accordance with  Lieb's theorem, but also inhomogeneous, with either
$U_p=0,\ U_d\neq 0$ or $U_p\neq 0,\ U_d=0$.  In all cases, we set the
local orbital energies $\varepsilon_p=\varepsilon_d=0$ and global
chemical potential $\mu=0$.  With these choices, particle-hole symmetry
holds even in the inhomogeneous case, which yields half filling
$\rho=1$.

For a model in which all sites ${\bf r}$ and orbitals $\alpha$ have the same on-site $U$, the two ways of writing the interaction, $U n_{{\bf r}\uparrow}^{\alpha} n_{{\bf r}\downarrow}^{\alpha} $ and $U (n_{{\bf r}\uparrow}^{\alpha}-\frac{1}{2} ) ( n_{{\bf r}\downarrow}^{\alpha}-\frac{1}{2}) $ differ only by a shift in the choice of the zero of
{\it global} chemical potential, so the physics is completely identical.
However, if $U_\alpha$ depends on $\alpha$ (or ${\bf r}$), then changing to a particle-hole symmetric form corresponds to an {\it orbital dependent} shift, i.e.~an unequal change in the individual $\varepsilon_\alpha$.
The symmetric form corresponds to a special choice in which the occupancies of all orbitals are identically half-filled.
Typically this choice is not obeyed in a real material, where each orbital has a unique filling.
However, since it is a prequisite for the applicability of Lieb's theorem, we impose it here.

The magnetic behavior is characterized by the local moments 
\begin{align}
\langle m^2_\alpha \rangle = 
\langle (n^\alpha_{{\bf r} \uparrow} - 
n^\alpha_{{\bf r} \downarrow})^2 \rangle 
\label{eq:moment}
\end{align}
and also by the real space spin-spin correlation functions 
\begin{align}
c^{\alpha \beta}({\bf r}) = 
\langle 
c^{\alpha \dagger}_{{\bf r_0}+{\bf r} \, \downarrow} 
c^{\alpha \phantom{\dagger}}_{{\bf r_0}+{\bf r} \, \uparrow} 
c^{\beta \dagger}_{{\bf r_0} \, \uparrow} 
c^{\beta \phantom{\dagger}}_{{\bf r_0} \, \downarrow} 
\rangle
\label{eq:Cspin}
\end{align}
which measure the result of raising a spin on site ${\bf r_0}$ in 
orbital $\beta$
and its subsequent lowering at site ${\bf r_0}+{\bf r}$ in orbital
$\alpha$.  The Fourier transforms of $c^{\alpha \beta}({\bf r})$ are the
magnetic structure factors,
\begin{align}
S^{\alpha \beta}({\bf q}) = \sum_{\bf r}
c^{\alpha \beta}({\bf r}) \, e^{i {\bf q} \cdot {\bf r}}
\label{eq:Sq}
\end{align}
In the ferrimagnetic state proposed by Lieb, $c^{\alpha \beta}({\bf
r})>0$, when $\alpha$ and $\beta$ are both $d$-, or both $p$-orbitals,
while for un-like orbitals $c^{\alpha \beta}({\bf r})<0$.  We will focus
on FM, ${\bf q}=0$.

In order to probe the metallic or insulating character of the system, a
useful quantity is the electronic compressibility, defined as
\begin{equation}
	\kappa = -\frac{1}{\rho^2}\frac{\partial \rho}{\partial \mu},
\end{equation}
where $\rho$ is the electronic density.
The properties of the Hamiltonian Eq.\,\eqref{eq:ham} will be solved
using determinant Quantum Monte Carlo
(DQMC).\cite{blankenbecler81,white89,dosSantos03b} This method provides
an exact solution, on real-space lattices of finite size, subject to
statistical error bars and (small) `Trotter errors' from the
discretization $\Delta \tau$ of imaginary time (inverse temperature).
We have chosen $\Delta \tau$ small enough so that these Trotter errors
are comparable to, or less than, the statistical errors on $c^{\alpha
\beta}({\bf r})$ and $S^{\alpha \beta}({\bf q})$.  We define the lattice
spacing ($a=1$) as the distance between nearest $d$-sites; accordingly,
the finite size $L$ (in units of lattice spacing) is given by the number
of $d$-sites along one direction, while the numerical effort is actually
measured by the number of lattice sites, $N_s\equiv 3(L\times L)$.    
Lattice separations along the horizontal or vertical directions
with $|{\bf r}|$ integer correspond to correlations between like
orbitals, whereas half-integral $|{\bf r}|$ denote unlike orbitals.

\begin{figure}[t]
\includegraphics[scale=0.36]{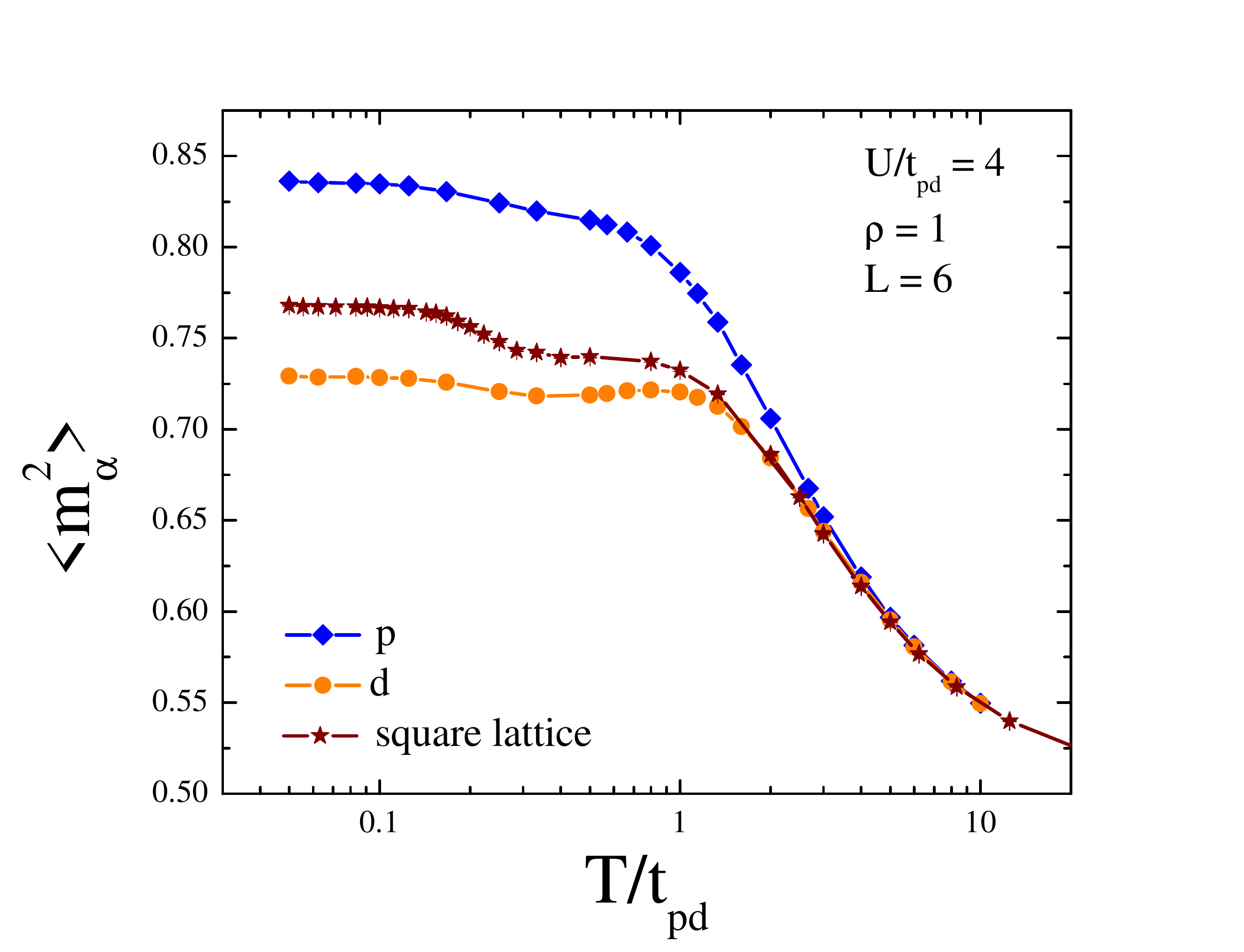} %{figure=momentU4.ps,width=2.7in,height=3.3in,angle=-90,clip}
\caption{(Color online) 
Temperature evolution of the local moment on $d$- and $p$-sites of the
Lieb lattice (linear-log scale).  The four-fold coordinated $d$-sites
have a lower moment than the two-fold coordinated $p$-sites.
Moment formation occurs mainly when $T/t_{pd} \sim U$, but a smaller
signal is also seen at $T/t_{pd}\sim J$, the exchange energy.
Here, and in all subsequent figures, when not shown, error bars 
are smaller than symbol size.
}
\label{fig:momentU4} % Fig2
\end{figure}

\begin{figure}[t]
\includegraphics[scale=0.35]{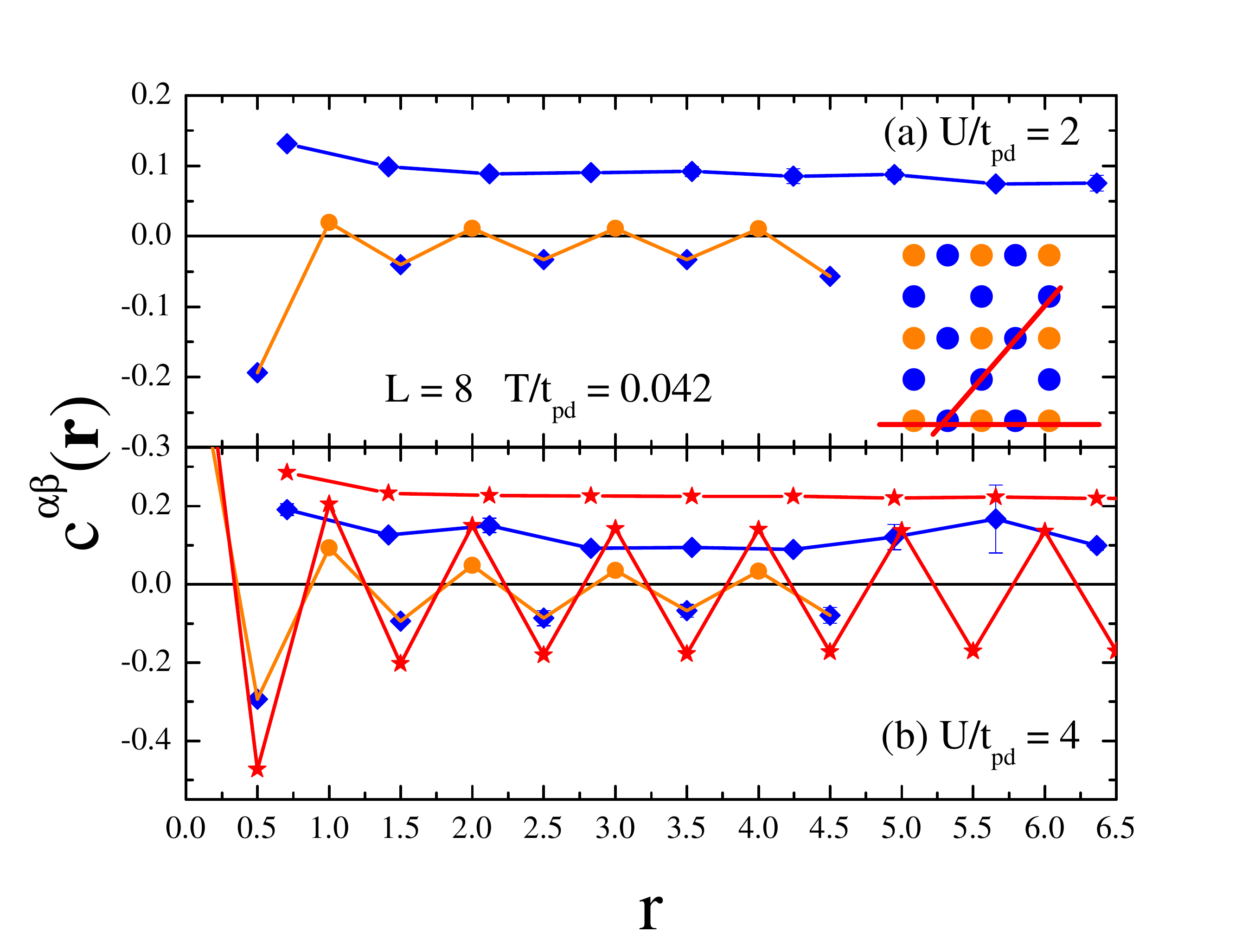} 
\caption{(Color online) Spatial dependence of spin-spin correlation
functions, Eq.\,\eqref{eq:Cspin}, at a  fixed temperature, for
$U/t_{pd}=2$ (top panel) and $U/t_{pd}=4$ (bottom panel).  
In each panel, diamonds and circles respectively represent positions of $p$- and $d$ sites, while stars on the bottom panel are data for the Heisenberg model. 
Curves going solely through diamonds (blue curves) correspond to placing the origin 
at a $p$ site, and $\mathbf{r}$ running over $p$ sites along a straight line
at an angle of $45^\circ$ (see the inset); 
curves alternating between diamonds and circles correspond to placing the origin 
on a $d$ site, and $\mathbf{r}$ running along a horizontal
line (see the inset).
}
\label{fig:cofr} % Fig 3
\end{figure}

\begin{figure}[h!]
\includegraphics[scale=0.35]{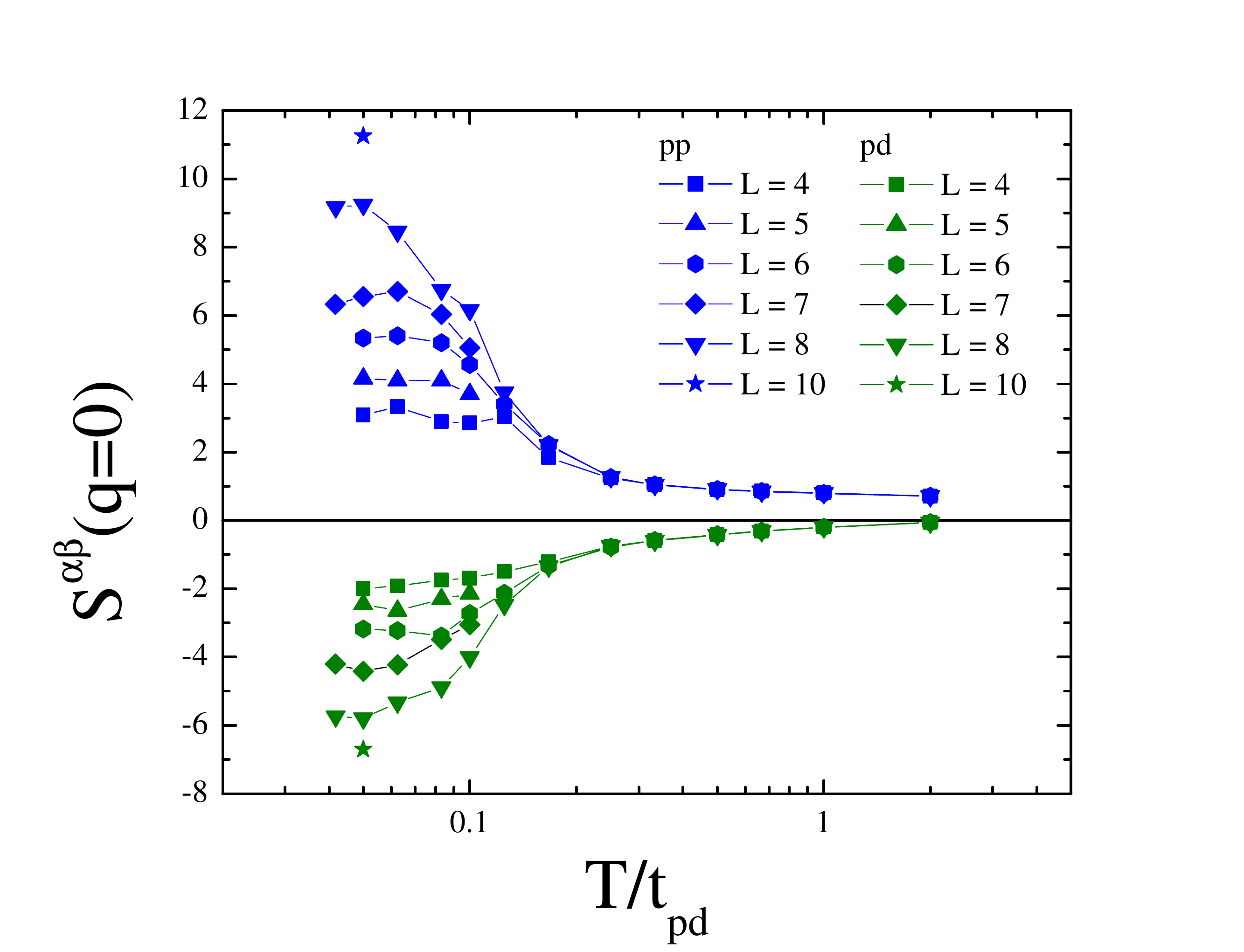} 
\caption{(Color online) 
The FM structure factors
$S^{\alpha\beta}({\bf q}=0,0)$ of the Lieb lattice are
plotted as functions of temperature for different lattice sizes $L$.
for $U/t_{pd}=4$.
The relative signs, 
$S^{p^x\!,p^x}\!\left[{\bf q}=(0,0)\right]>0$ 
and
$S^{d,p^x}\!\left[{\bf q}=(0,0)\right]<0$,
are signatures of ferrimagnetism.
At high $T$, where the real space
correlations are short range,
$S^{\alpha\beta}$ is independent of $L$.  As $T$ decreases,
$S^{\alpha\beta}$ plateaus at successively larger values for 
increasing $L$, providing evidence that spin correlations extend over
the entire lattice.
}
\label{fig:SnallU4q0thinned}  %Fig 4
\end{figure}

\begin{figure}[t]
\includegraphics[scale=0.36]{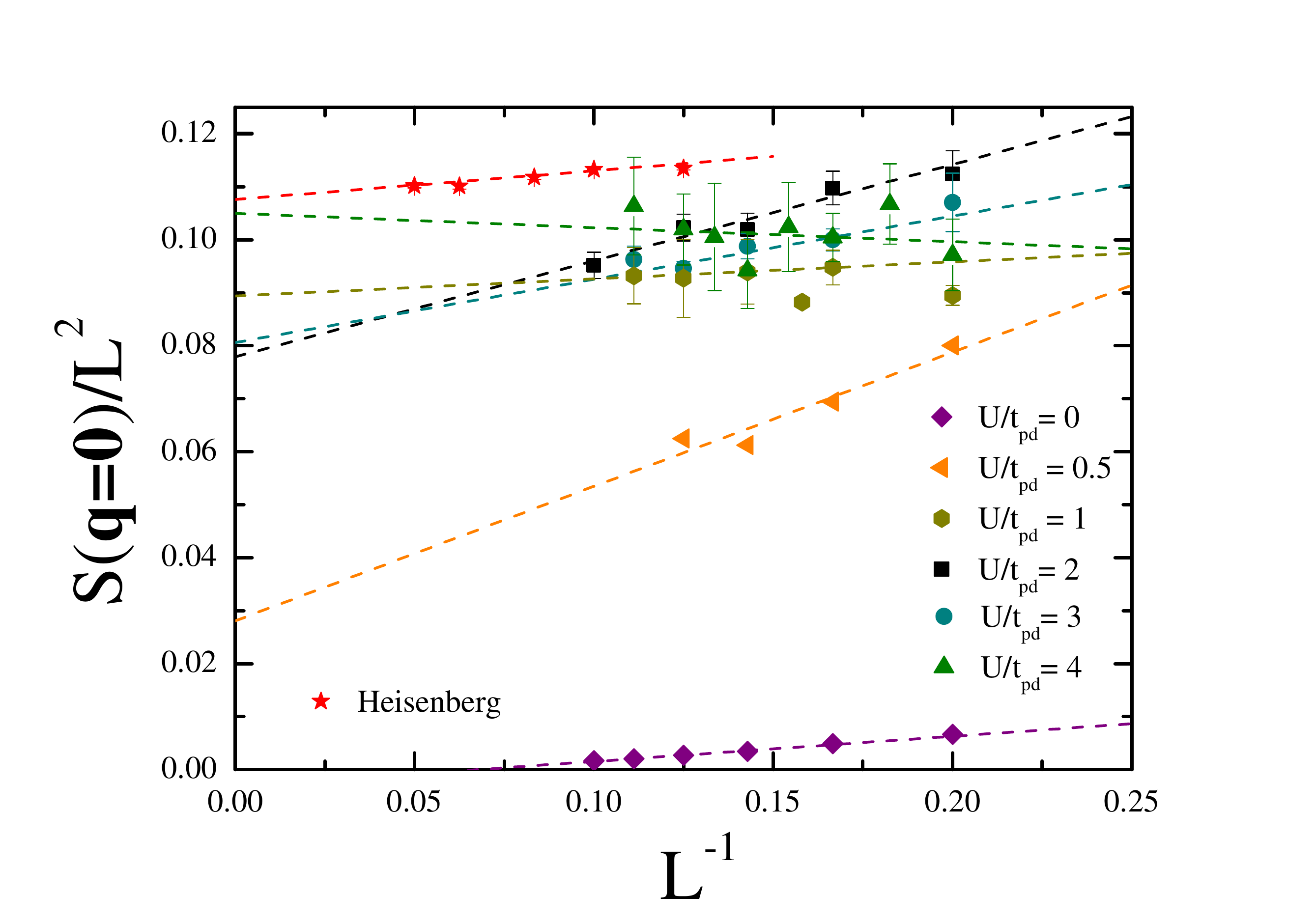} 
\caption{(Color online) Finite-size scaling plots for the normalized
ground state structure factor $m_{\rm F}^2$. For each $U\neq0$, they
extrapolate to a non-zero value in the thermodynamic limit.  The data
labelled Heisenberg have been obtained for localized spins on Lieb
lattice, interacting through nearest-neighbor exchange coupling
$\mathbf{S}_i\cdot\mathbf{S}_j$; see text.  
}
\label{fig:S_global-FSS} %Fig 5
\end{figure}

\begin{figure}[h!]
\includegraphics[scale=0.35]{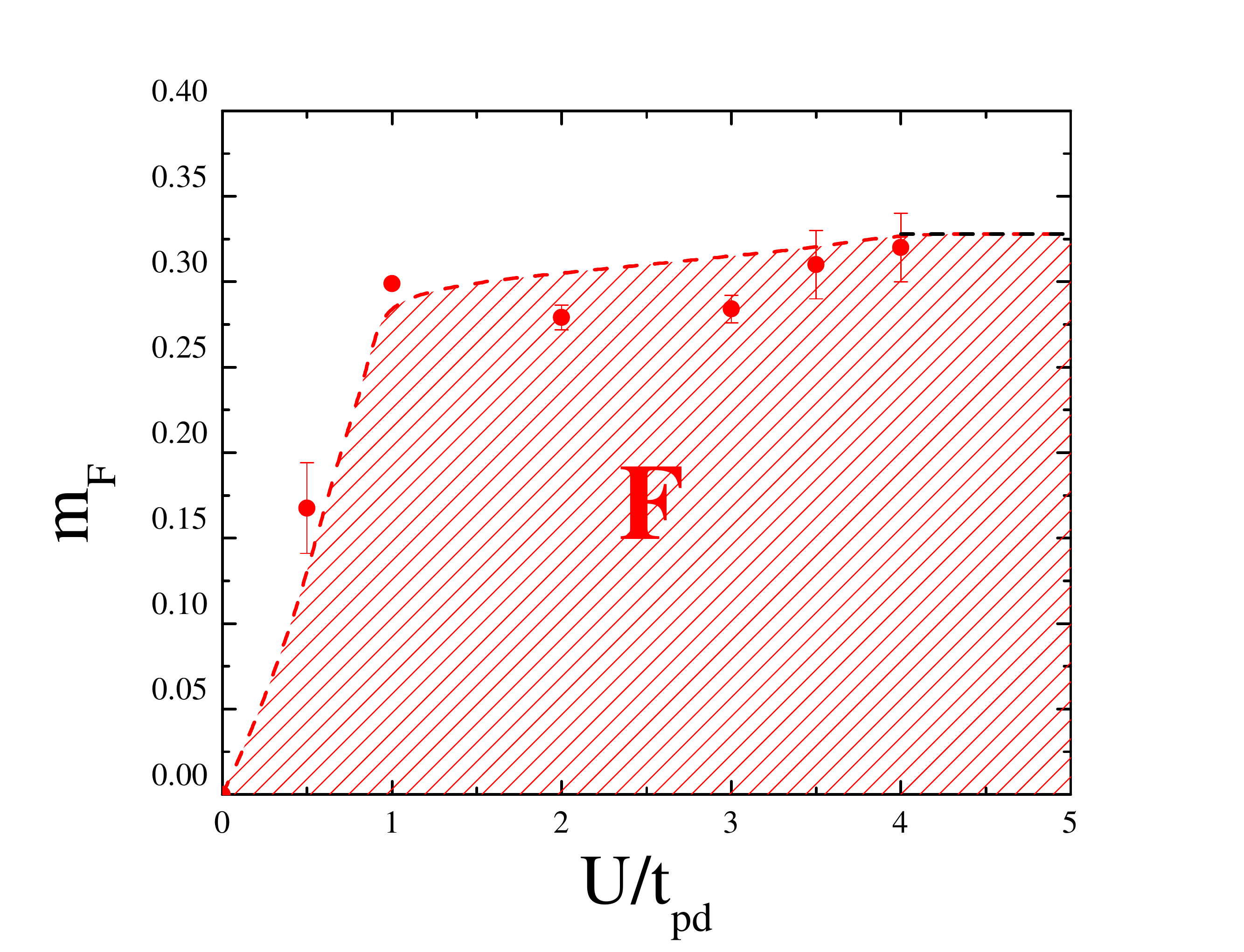} 
\caption{(Color online) Global ferromagnetic order parameter as a
function of the on-site repulsion, $U$, obtained from the extrapolated
values. % (intercepts with the vertical axis) in
Fig.\,\ref{fig:S_global-FSS}.  The (red) dashed line going through the
data points is a guide to the eye, while the horizontal (black) dashed
line is the Heisenberg limit. 
}
\label{fig:m_F-vs-U} %Fig 6
\end{figure}

%%%%%%%%%%%%%%%%%%%%%%%%%%%%%%%%%%%%%%%%%%%%%%%%%%%%%%%%%%%%%%%%%%
%%%%%%%%%%%%%%%%%%%%%%%%%%%%%%%%%%%%%%%%%%%%%%%%%%%%%%%%%%%%%%%%%%
\section{The Homogeneous Lattice}
\label{sec:homog}
%%%%%%%%%%%%%%%%%%%%%%%%%%%%%%%%%%%%%%%%%%%%%%%%%%%%%%%%%%%%%%%%%%
%%%%%%%%%%%%%%%%%%%%%%%%%%%%%%%%%%%%%%%%%%%%%%%%%%%%%%%%%%%%%%%%%%

Figure \ref{fig:momentU4} shows the temperature evolution of the local
moment on $d$ and $p$ sites.  Both moments start at the common high
temperature value $\langle m_\alpha^2 \rangle = \frac{1}{2}$ and become
better formed as the temperature crosses the energy scale $T \sim U$.
At low temperatures, the moments stabilise in plateaux with $\langle
m_\alpha^2 \rangle <1 $, which reflect residual quantum fluctuations
arising from $t_{pd}/U \neq0 $.  These fluctuations are larger for the
$d$-sites, which have four neighboring $p$-sites, than for the $p$-sites
which have only two neighboring $d$-sites.  It is also interesting to
note that the local moment for the usual square lattice, $\langle
m_{\mathrm{square}}^2\rangle$, is such that $\langle m_{d}^2\rangle <
\langle m_{\mathrm{square}}^2\rangle < \langle m_{p^{x(y)}}^2\rangle$.

Inter-site spin correlations develop at lower temperatures associated
with the exchange energy scale $J \sim t_{pd}^2/U$.  Figure
\ref{fig:cofr} illustrates the different behaviors of correlations with
the distance (all consistent with the rigorous results for their signs, as
derived in Ref.\,\onlinecite{shen94}), at a fixed low temperature,
$T/t_{pd} = 0.042$.  
Along a path which only includes $p$ sites [(blue) curve going solely through diamond data points], correlations are always positive,  indicating a ferromagnetic alignment, and with a robust persistence at large
distances.  By contrast, along a horizontal path which includes both $d$
and $p$ sites, the correlations alternate in sign, consistently with AFM
alignment between $d$ and $p$ sites, and a FM alignment between $d$
sites; here again, the persistence of correlations at large distances
($\sim L/2$) suggests an overall long-range FM order.  
Also shown in Fig.\,\ref{fig:cofr}(b) are data for the Heisenberg model on the same lattice, which corresponds to the strong coupling limit ($U\gg t_{pd}$) of the Hubbard model; these latter data have been obtained through the stochastic series expansions (SSE) method.\cite{Syljuasen02,Sandvik03}
The amplitudes for $U/t_{pd}=4$ are still quite far from their strong coupling limit, but one can infer that the slow decay of correlations is a dominant feature, which can therefore being taken as indicative of long range order in the ground state for all $U_{pd}$.

At high temperatures, $c^{\alpha \beta}({\bf r})$ is short ranged, so
the sum over all lattice sites in the structure factor is independent of
system size.  This is reflected in the high-temperature collapse of
$S^{\alpha \beta}({\bf q}=(0,0))$  in Fig.\,\ref{fig:SnallU4q0thinned}.
Data for $S^{\alpha \beta}({\bf q}=(0,0))$ for different $L$ split apart
at $T \sim J$.

A more rigorous probe of long range order is carried out through
finite-size scaling analyses.\cite{huse88}  The square of the order
parameter is obtained by normalizing the structure factor to the lattice
size, $m^2_{\mathbf{q}} = S^{\alpha\beta}(\mathbf{q})/L^2$.  This will have a
nonzero value in the thermodynamic limit $1/L \rightarrow 0$, if
$c^{\alpha\beta}({\bf r})$ is long-ranged, with a $1/L$ correction.  In
Figure \ref{fig:S_global-FSS} data for the global FM structure factor
are displayed, for several values of $U$;   
also shown are data for the Heisenberg model on the same lattice.
The extrapolated values of the order parameter are shown in Fig.\,\ref{fig:m_F-vs-U}, as a function of $U$, thus confirming the existence of long range ferromagnetic order for all $U>0$.
Note that $m_F$ rises sharply for $U/t_{pd} \lesssim 1$, and then stabilizes towards the Heisenberg model value for large $U/t_{pd}$. 
At this point, a technical remark is worth making: for $U/t_{pd}\gtrsim 4$ one has to perform simulations at very low temperatures ($T \lesssim 0.025 t_{pd}$, or $\beta\equiv t_{pd}/T=40$) 
in order to ensure the structure factor has stabilized; these temperatures are much lower than those needed for the simple square lattice with the same $U/t$,  $\beta\gtrsim 25$.\cite{Varney2009}

\begin{figure}[t]
\includegraphics[scale=0.36]{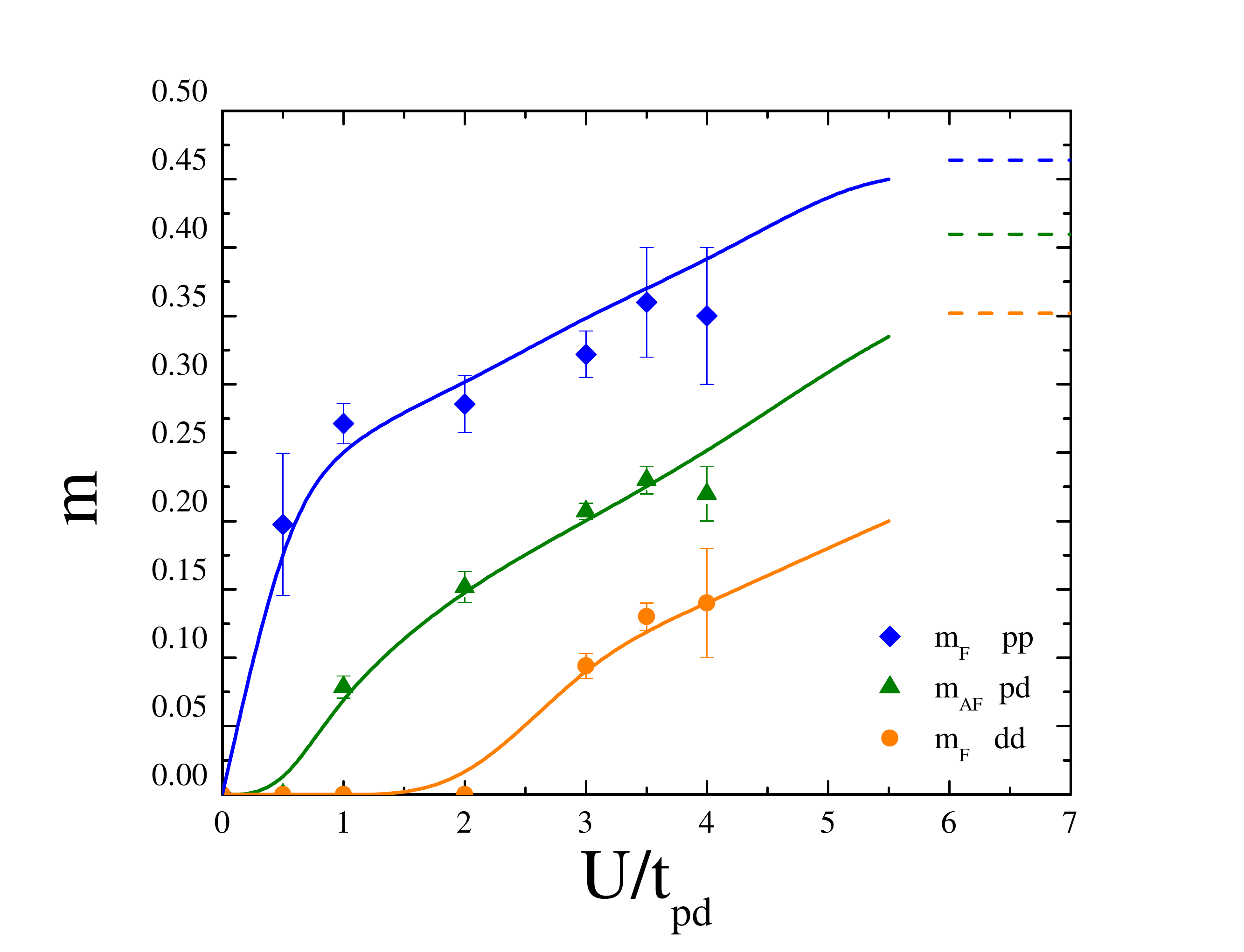} 
\caption{(Color online) Extrapolated ($L\to \infty$) values of the
channel-resolved order parameters obtained from the scaling of the
structure factor; see text.  For the FM channels ($dd$ and $pp$) we set
$\mathbf{q}=0$ in Eq.\,\eqref{eq:Sq}, while for the AFM channel ($dp$)
the sum is carried out with opposite signs at adjacent sites. %set
$\mathbf{q}=(\pi,\pi)$.
}
\label{fig:m_ab-vs-U} %Fig 7
\end{figure}

If we now perform separate finite-size scaling analyses for the
structure factors in the different channels, $S^{\alpha \beta}({\bf q})$
with $\alpha,\beta=d, p^x, p^y$, we can probe the corresponding
sub-lattice order parameters; their dependence with $U$ is shown in
Fig.\,\ref{fig:m_ab-vs-U}.  It is interesting to see that $pp$
ferromagnetism rises sharply with $U$, in marked contrast to the very
slow rise in the $dd$ sublattice.  A strong coupling analysis of the
$p^ydp^x$ cluster of three Heisenberg-coupled spins reveals that the two
$p$ spins form a triplet, which adds to the $d$ spin, leading to a total
spin $S_{\mathrm{cluster}}=1/2$ characterizing a
\textit{ferrimagnetic} state; this picture can also be applied in weak
coupling, as a result of the flat $p$-band.  We may therefore attribute
the sharper rise of the $pp$ FM order parameter as due to the $p$ spins
locking into triplets as soon as $U$ is switched on, while the $d$ spin
is somewhat shielded by the surrounding triplets.  Figure
\ref{fig:m_ab-vs-U} also shows that the data converge very slowly to the
Heisenberg limit; again this may be attributed to the difference in the
number of nearest neighbors of $p$ and $d$ sites.

%%%%%%%%%%%%%%%%%%%%%%%%%%%%%%%%%%%%%%%%%%%%%%%%%%%%%%%%%%%%%%%%%%
%%%%%%%%%%%%%%%%%%%%%%%%%%%%%%%%%%%%%%%%%%%%%%%%%%%%%%%%%%%%%%%%%%
%\section{Spectral Function}
%%%%%%%%%%%%%%%%%%%%%%%%%%%%%%%%%%%%%%%%%%%%%%%%%%%%%%%%%%%%%%%%%%
%%%%%%%%%%%%%%%%%%%%%%%%%%%%%%%%%%%%%%%%%%%%%%%%%%%%%%%%%%%%%%%%%%
It is also worth checking the insulating nature of the ferrimagnetic state.
To this end, we calculate the density of states $N(\omega)$ from DQMC data for the imaginary-time dependent Green's function, which is achieved by inverting the integral
equation,
\begin{align}
G(\tau) = \int \mathrm{d}\omega\, N(\omega) \,
\frac{e^{-\omega \tau}}{e^{\beta \omega} + 1}.
\label{eq:aw}
\end{align}
This inversion can be done, for example, with the `maximum entropy'
method.\cite{jarrell96} In the case of the square lattice, $N(\omega)$
exhibits a gap at half-filling;\cite{white89}  this `Slater' gap
originates in AFM order at weak $U$ and crosses over into a Mott gap at
strong coupling.

\begin{figure}[t]
\includegraphics[scale=0.35]{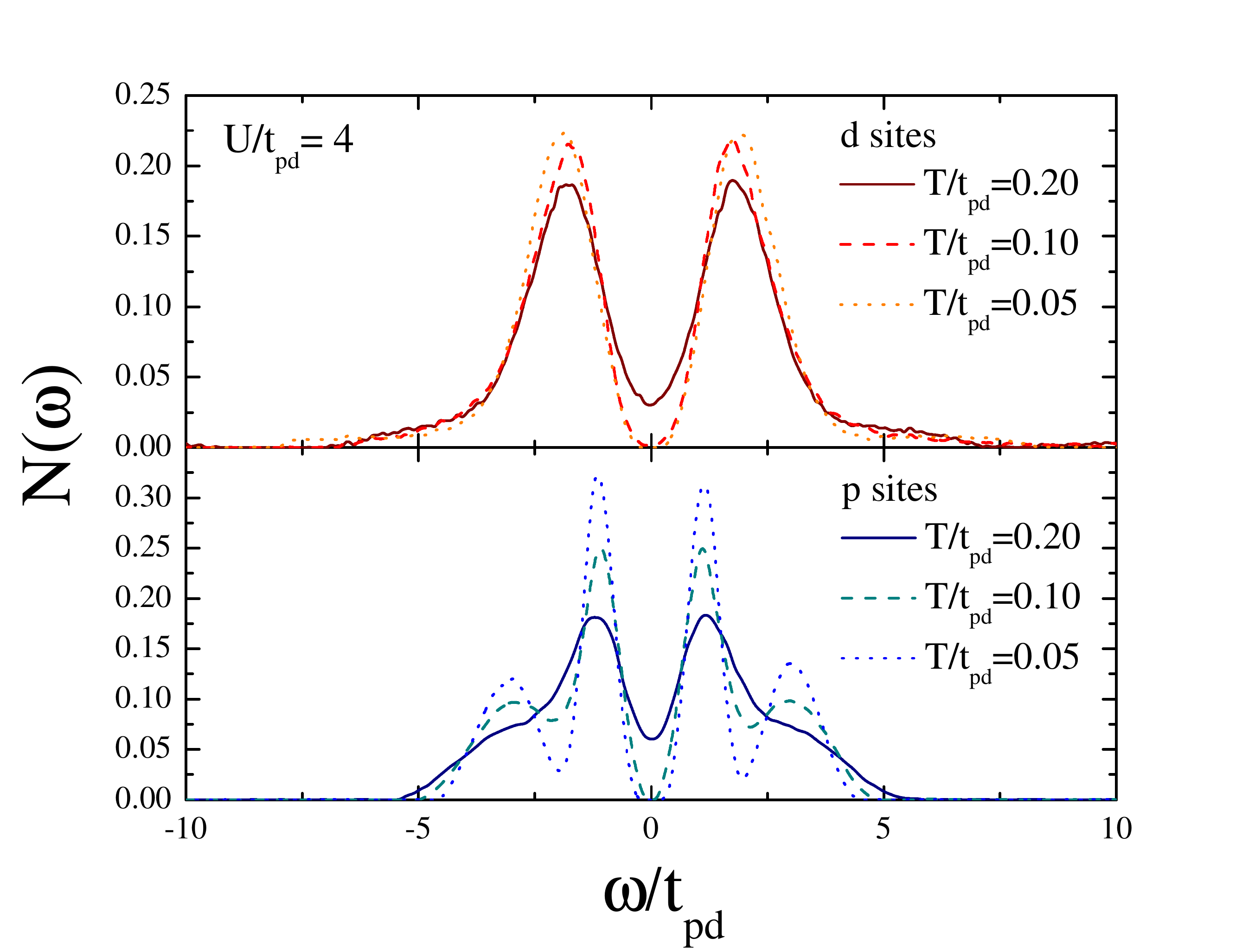} 
\caption{(Color online) 
Local density of states on $d$ sites (top panel) and on $p$ sites (bottom panel), at three different temperatures.
}
\label{fig:DOS} %Fig 8
\end{figure}

\begin{figure}[t]
\includegraphics[scale=0.375]{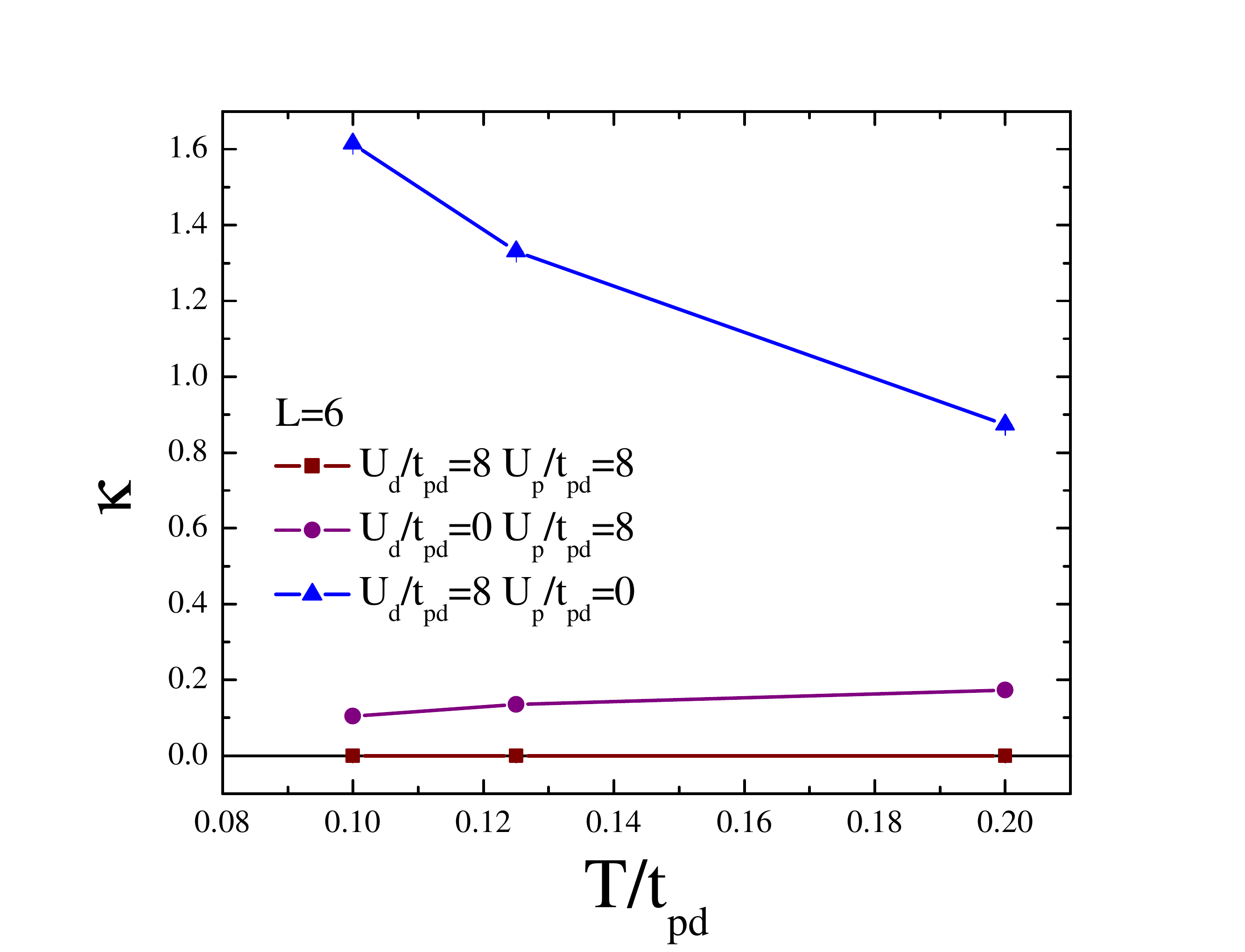} 
\caption{(Color online) 
Comparison of the compressibility at half filling in three instances:
homogeneous lattice (squares), $U_d=0$ (circles), and $U_p=0$
(triangles). When non-zero, the $U$'s are all set to $8t_{pd}$; the
linear lattice size is $L=6$.
}
\label{fig:kappa} %Fig 9
\end{figure}

Figure \ref{fig:DOS} shows the projected density of states for the Lieb
lattice.  We see that at sufficiently low temperatures an insulating gap
develops for both orbitals, similarly to the square lattice,
but with the important difference that in the present case it results
from a ferromagnetic state.  Further, the density of states on the $p$
sites displays a double-peak structure on each side of the Fermi energy.
The additional peaks originate from the splitting of the flat band on
the $p$ sites when a ferromagnetic state is formed; this is similar to
what happens in the periodic Anderson model  when the Kondo resonance is
split when an antiferromagnetic state is formed.  We have also obtained
the density of states for other values of $U$. 
The gap increases monotonically with $U$.

\section{The Inhomogeneous lattices}
\label{sec:inhomog}

The strong coupling limit of a generic inhomogeneous Lieb lattice at
half filling (single occupancy enforced on every site), and with
$U_d\neq U_p,\ U_d, U_p>0$ corresponds to a Heisenberg
model with uniform exchange\cite{Franca10} 
\begin{equation}
	J'=\frac{4t^2}{\widetilde{U}} 
\label{eq:Jayp}
\end{equation}
where $\widetilde{U}$ is the geometric mean between the on-site repulsion on adjacent sites,
\begin{equation}
	\widetilde{U}=\frac{2 U_pU_d}{U_p+U_d}.
\label{eq:Utilde}
\end{equation}
Since one of the steps in Lieb's proof relies on the strong coupling limit of
the Hubbard model,\cite{lieb89,lieb62}  the existence of a ferromagnetic state
also holds in this case.  
This is discussed further in the conclusions.

However, if either $U_d$ or $U_p$ vanishes, this correspondence with the Heisenberg model completely breaks down -- single occupancy on every site is no longer guaranteed even at half filling.  
Further, due to the different
neighborhoods of the $d$ sites (4 $p$ neighbors) and of the $p$ sites (2
$d$ neighbors), switching off $U_d$ or $U_p$ leads to radically
different effects, as we now discuss.  Figure \ref{fig:kappa} displays
data for the compressibility.  We see that when $U_d=0$ the system
behaves as an insulator; by contrast, when $U_p=0$ the compressibility
increases as the temperature decreases, indicating a metallic state.
Therefore, when $U_d=0$ and at half filling, each $p$ site is occupied
by one fermion, so that the $d$ site is also singly occupied, as if
$U_d$ were non-zero; from the magnetic point of view, one then expects a
ferromagnetic ground state, just as in the homogeneous case.  When
$U_p=0$ the likelihood of double occupancy of the $p$ sites increases,
thus destroying any magnetic ordering. As we will see, these
expectations are borne out by our simulations. 

\begin{figure}[t]
\includegraphics[scale=0.35]{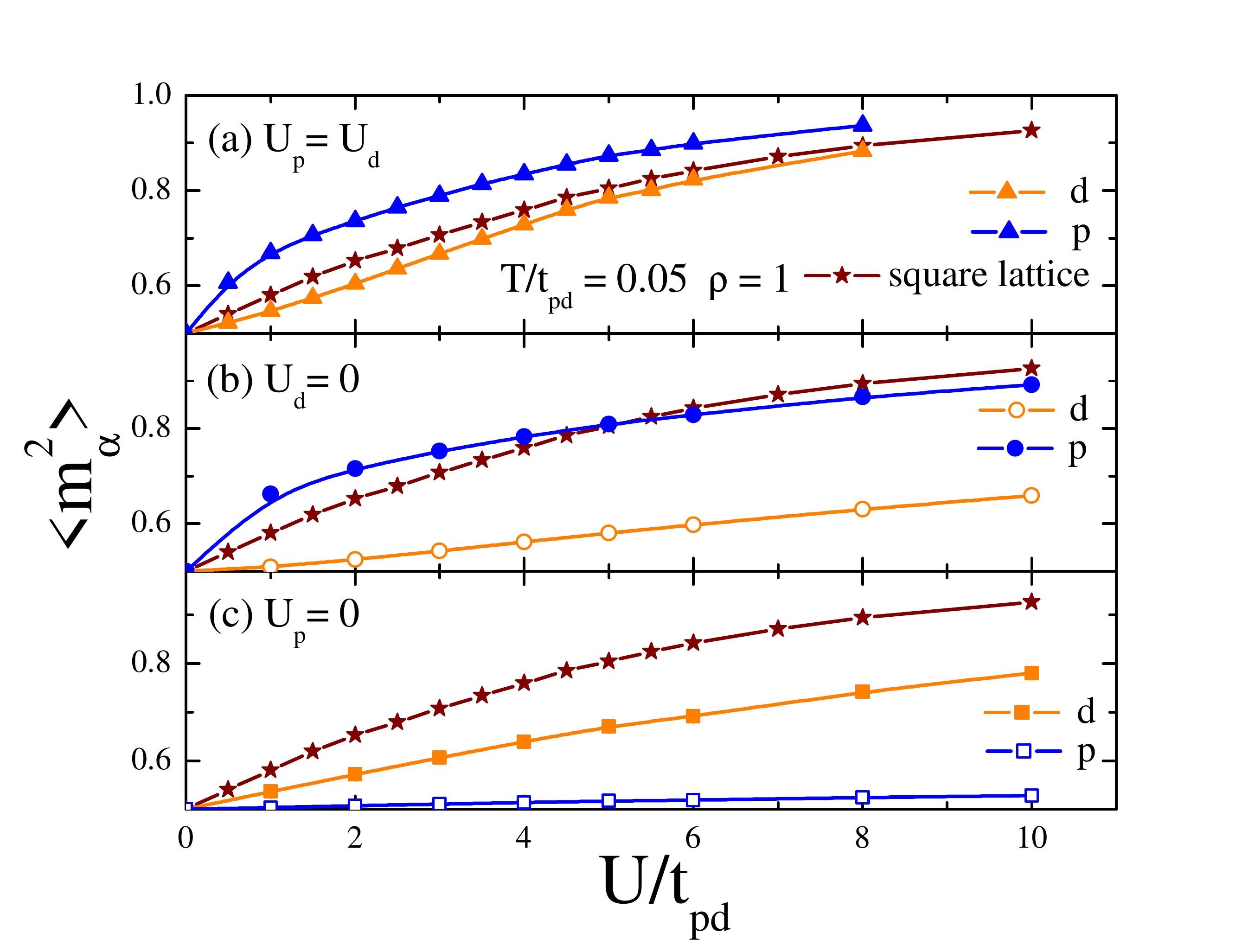} 
\caption{(Color online) 
Local moment on $d$ sites [light (orange) color] and on $p$ sites [dark
(blue) color], as functions of the on-site repulsion for (a) the
homogeneous case, (b) $U_p=U,\ U_d=0$, and (c) $U_d=U,\ U_p=0$. Data for
the usual square lattice are also shown (stars), for comparison. 
}
\label{fig:s2nonLieb} %Fig 10
\end{figure}

Figure \ref{fig:s2nonLieb} compares the local moment in the homogeneous
and inhomogeneous `Lieb lattices'.  One immediate effect of switching
off the repulsion on a subset of sites is the strong suppression of the
local moment on exactly those `free' sites; this suppression is almost
complete (becoming very near the minimum value of 1/2) on $p$ sites when
$U_p=0$.  However, when $U_d=0$, and $U_p\lesssim 4t_{pd}$ the local
moment on the $p$ sites is not significantly affected in comparison with
the homogeneous case; for $U\gtrsim 4 t_{pd}$ it becomes slightly
smaller than the one for the square lattice.  By contrast, when $U_p=0$
the suppression of $\langle m^2\rangle$ on the $d$ sites takes place for
all $U$, as a result of increasing double occupancy.

\begin{figure}[t]
\includegraphics[scale=0.35]{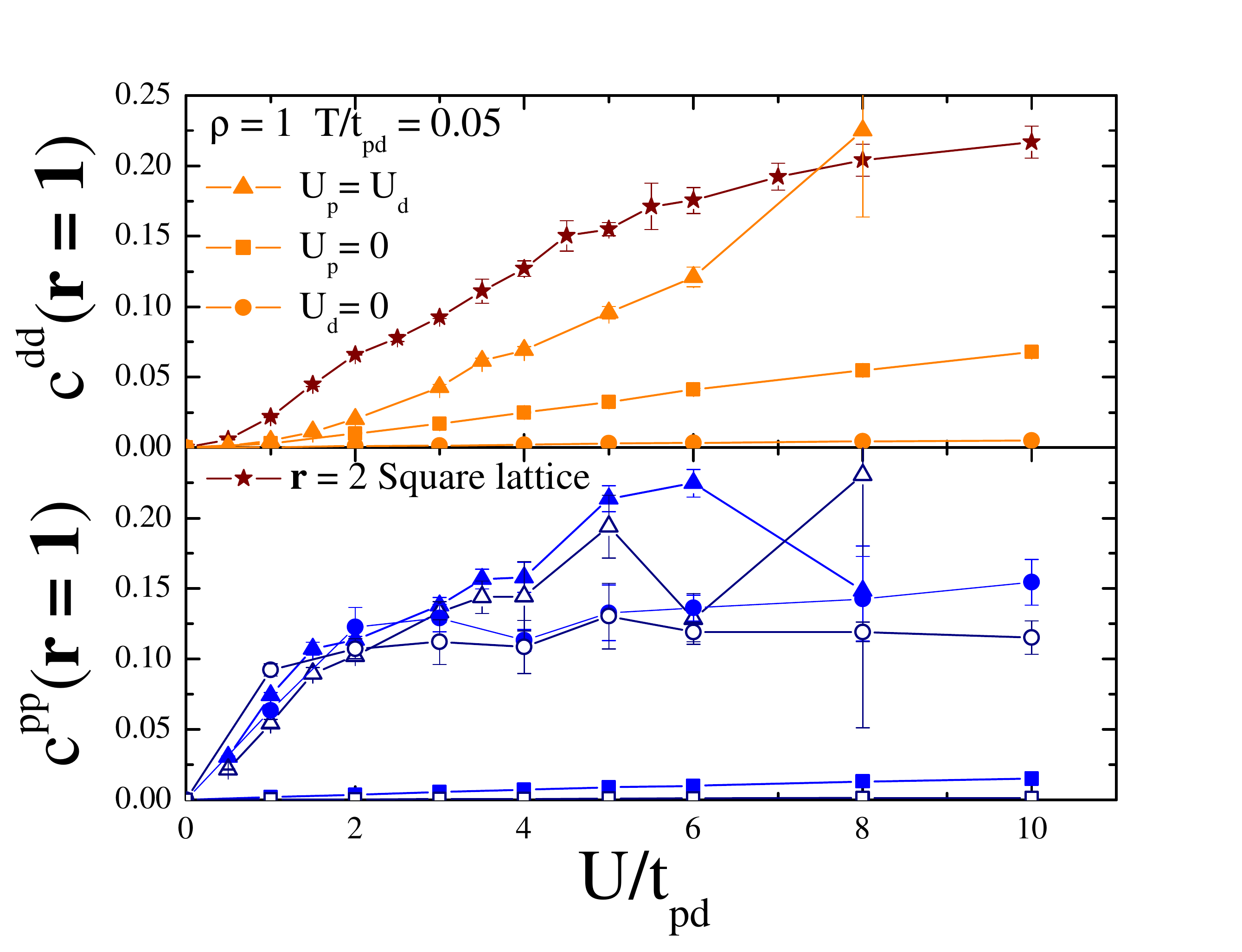} 
\caption{(Color online) 
Spin correlations between first neighbor like-sites. The upper panel
displays the correlations between $d$ sites, while the lower panel shows
those between $p$ sites; in the latter case, the $p$ sites may have an
intervening $d$ site (filled symbols), or not (empty symbols). 
Data for the usual square lattice are shown, but
the fair comparison in this case is with sites on different sublattices
(hence two lattice spacings apart, or $r=2$); see text.
}
\label{fig:c1nonLieb} %Fig 11
\end{figure}

In Fig.\,\ref{fig:c1nonLieb} we show the spin correlation between sites
one lattice spacing apart, as functions of the on-site repulsion.  The
$dd$ correlations for the homogeneous Lieb lattice are suppressed in
comparison with those for the square lattice.  As noted earlier in
connection with Fig.\,\ref{fig:momentU4}, the local moment on the $d$
site is smaller than on the square lattice, so this reduction 
in correlations between
spins on $d$ sites is expected.  For the
inhomogeneous Lieb lattice, correlations between spins on $d$ sites are
suppressed even more, with those on the $d$ sites being completely
suppressed when $U_d=0$.  By contrast, the $pp$ correlations are quite
robust if the Coulomb repulsion is only switched off on the $d$ sites;
when $U_p=0$, $pp$ correlations are strongly suppressed.  As
anticipated, repulsion on the $p$ sites is crucial to the onset of
ferromagnetic correlations.
At this point, a comment should be made: in the strong coupling (i.e., Heisenberg) limit, $pp$ correlations one lattice spacing apart are exactly the same irrespective of including, or not, an intervening $d$-site. However, up to the couplings covered in Fig.\,\ref{fig:c1nonLieb}, the noticeable difference is due to both the temperature not being low enough, and to the coupling being not so strong.   

This is even more evident when we probe long range order (LRO) through
finite-size scaling analyses of the %channel-resolved structure factor,
$m_{\alpha,\beta}^2=S^{\alpha,\beta}/L^2$.  When $U_p=0$ the overall
ferromagnetic order parameter decreases very fast as $L\to\infty$,
indicating the absence of LRO.  This is reminiscent of what happens in
the diluted Hubbard model on a square lattice.  LRO in the ground state
is only possible below a certain threshold $f_c$ of free sites, which
depends on the strength of the on-site
interaction:\cite{Ulmke98b,Hurt05,Mondaini08} here an effective fraction
of free sites can be taken as $f=2/3$, which is above the thresholds
$f_c^\mathrm{square}(U=8t)\simeq 0.4$, and for
$f_c^\mathrm{square}(U=-4t)\simeq 0.3$.  This is consistent with
previous work on the case $U_p=0$
in models of CuO$_2$ sheets of cuprate superconductors, which do not
display an antiferromagnetic ground state, unless a site energy
difference $\varepsilon_p - \varepsilon_d > 0$ is
included to enhance charge disproportionation.\cite{Scalettar91,Kung16}

\begin{figure}[t]
\includegraphics[scale=0.36]{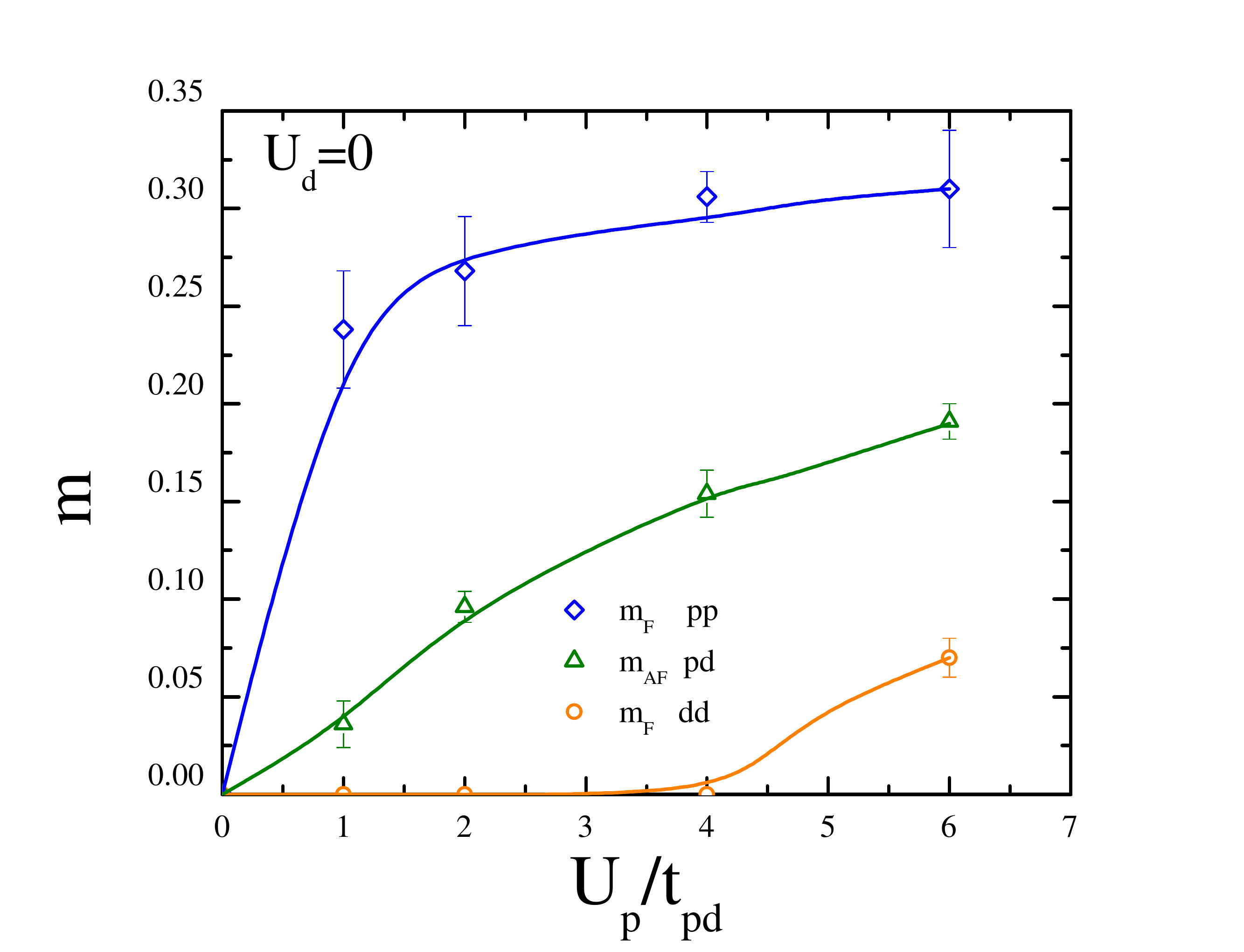} 
\caption{(Color online) 
Same as Fig.\,\ref{fig:m_ab-vs-U}, but now for the $U_d=0$ case. 
}
\label{fig:m_ab-vs-Up} %Fig 12
\end{figure}

\begin{figure}[t]
\includegraphics[scale=0.35]{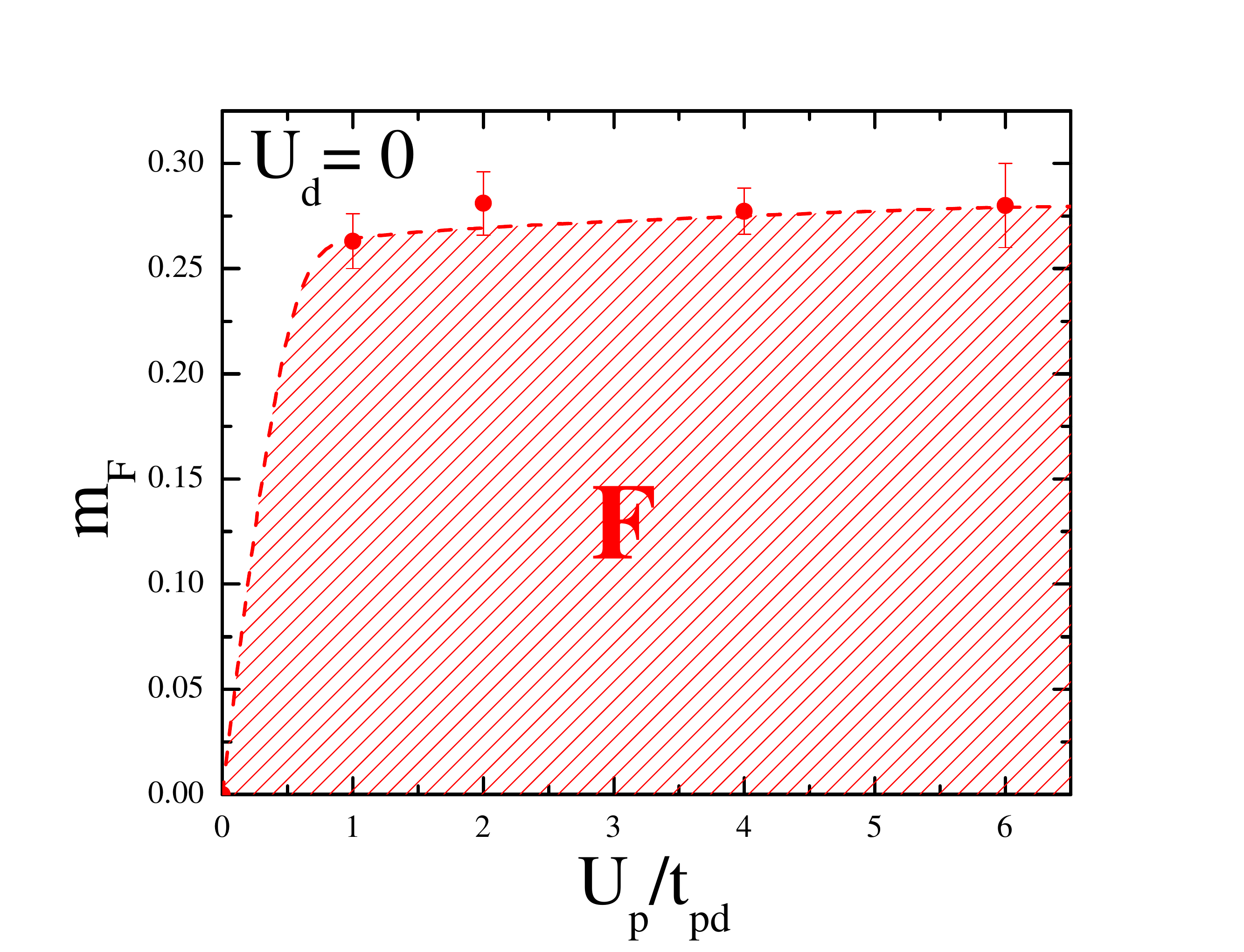} 
\caption{(Color online) 
Global ferromagnetic order parameter as a function of the on-site repulsion, $U_p/t_{pd}$, obtained from the extrapolated values for the case $U_d=0$. % (intercepts with the vertical axis) in Fig.\,\ref{fig:S_global-FSS}. 
The (red) dashed line going through the data points is a guide to the eye. 
}
\label{fig:m_F-vs-Up} %Fig 13
\end{figure}

For $U_d=0$, a finite-size scaling analysis of the overall ferromagnetic
order parameter indicates LRO.  The channel-resolved extrapolated order
parameters shown in Fig.\,\ref{fig:m_ab-vs-Up} are very similar to those
for the homogeneous case; the same is true for the global ferromagnetic
order parameter, as shown in Fig.\,\ref{fig:m_F-vs-Up}.  We therefore
conclude that the pre-conditions for ferromagnetism on Lieb lattices, at
least as far as homogeneity is concerned, are less restrictive than
those originally assumed in Lieb's proof of the theorem.

\section{Away from half filling}
\label{sec:doped}

\begin{figure}[t!]
\includegraphics[scale=0.35]{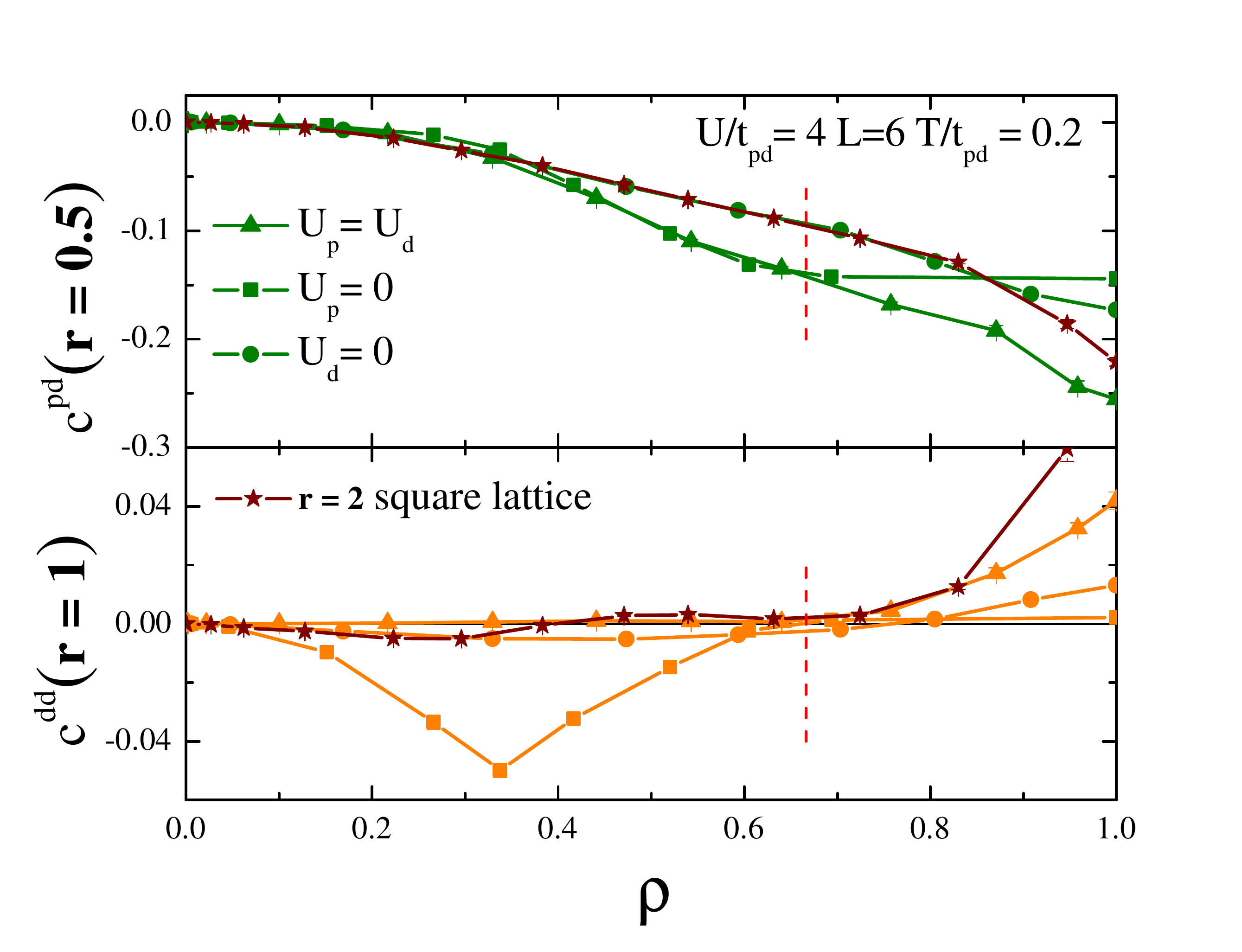} 
\caption{(Color online) 
Spin correlation functions as functions of band filling at fixed temperature $T=0.17t_{pd}$, for a lattice with $6\times 6$ $ppd$ cells, highlighting the differences between homogeneous and both inhomogeneous cases.  
Top panel: correlations between spins on nearest $s$ and $p$ sites. 
Bottom panel: correlations between spins on nearest $d$ sites.
}
\label{fig:doped} %Fig 14
\end{figure}

Away from half filling, the `minus-sign problem' (see, e.g.,
Refs.\,\onlinecite{Loh90,dosSantos03b}) hinders a thorough analysis at
low temperatures.  Nonetheless, some interesting conclusions may be
drawn at accessible temperatures (down to $T/t_{pd}=0.17$, or $T/W=0.03$
in units of the noninteracting bandwidth $W=4\sqrt{2}\,t$).  Figure
\ref{fig:doped}(top) shows that correlations between spins on near
neighbor $p$ and $d$ sites, $c^{pd}(r=0.5)$, are always AF (negative)
and increase monotonically in magnitude 
with $\rho$, up to half-filling $\rho=1$,
in both the H and IH cases.  The correlations between pairs of $d$
sites, $c^{dd}(r=1)$, Figure \ref{fig:doped}(bottom), show a more
intriguing behavior.  $c^{dd}(r=1)$ is small except near half-filling
where it turns relatively strongly positive for the H case (though less
large than on a square lattice), and weakly positive for the IH case
with $U_d=0$.

The richest structure is exhibited by $c^{dd}(r=1)$ for the IH case with
$U_p=0$.   It is largest in absolute value at filling
$\rho=1/3$, one fermion on each $d$ site,
while the $p$ sites are left empty.  This corresponds rather closely to
the situation of the CuO$_2$ planes in cuprates where $U_d > U_p$ and
the parent compound La$_2$CuO$_4$ has one hole per copper atom.   In the
cuprates, the site hole energy difference $\varepsilon_p - \varepsilon_d$ is
substantial, confining the holes to the copper sites, which
would enhance AF order further.  The importamt message of
Fig.\,\ref{fig:doped} (bottom) is that even in the absence of a
substantial $\varepsilon_p - \varepsilon_d$ there is robust (local) AF order.

%%%%%%%%%%%%%%%%%%%%%%%%%%%%%%%%%%%%%%%%%%%%%%%%%%%%%%%%%%%%%%%%%%
%%%%%%%%%%%%%%%%%%%%%%%%%%%%%%%%%%%%%%%%%%%%%%%%%%%%%%%%%%%%%%%%%%
\section{Conclusions}
\label{sec:conc}
%%%%%%%%%%%%%%%%%%%%%%%%%%%%%%%%%%%%%%%%%%%%%%%%%%%%%%%%%%%%%%%%%%
%%%%%%%%%%%%%%%%%%%%%%%%%%%%%%%%%%%%%%%%%%%%%%%%%%%%%%%%%%%%%%%%%%

As with the Anderson localization problem, where two dimensions occupies
a special position, itinerant ferromagnetism in 2D lies poised between
the 1D case where it is explicitly forbidden\cite{lieb62} and 3D where
it is (fairly) commonly observed in nature;
bounds on correlation functions for the Hubbard model (and some variants) in 
one- and two dimensions rule out any magnetic ordering at finite 
temperatures.\cite{Koma1992}
One route to ferromagnetism
was devised by Lieb,\cite{lieb89} who proved that the half-filled
Hubbard model on a bipartite lattice with unequal number of sites on
each sublattice has a non-zero total spin.  A particular geometry to
which this applies is the `decorated square lattice', also known as
`CuO$_2$ lattice', or `Lieb lattice': $d$ sites on the vertices of a
square lattice have $p$ sites as nearest neighbors at the mid-points between
the $d$ sites; see Fig.\,\ref{fig:CuO2}.  In this paper, we have used
Quantum Monte Carlo to unveil several details about the Lieb lattice, by
considering both the homogeneous case (on-site repulsion, $U$, has the
same magnitude on every site), as well as inhomogeneous cases, switching
off $U$ on either $p$ or $d$ sites.

For the homogeneous case, we have established that the magnitude of the
local moment is strongly dependent on the environment, being larger on
the $p$ sites than on the $d$ sites: fewer neighbors leads to a decrease
in itinerancy.  By analyzing the spatial decay of spin correlation
functions, and the lattice-size dependence of magnetic structure factor,
we have also provided numerical evidence for the existence of long range
ferromagnetic (or, \emph{ferrimagnetic}) order.  Interestingly, the
breakup into sublattice order parameters reveals that the ferromagnetism
of spins on $p$ sites is the most intense in magnitude, followed by the
antiferromagnetism along the square lattice directions ($d$-$p$ sites),
with the ferromagnetism of $d$ sites being the weakest.  These combine
to yield an overall ferromagnetic order parameter displaying
a sharp rise in the region $U/t_{pd}\lesssim 1$, and stabilizing towards the Heisenberg limit for $U/t_{pd}\gg 1$.
Further, by examining the
projected density of states (obtained with the aid of the maximum
entropy method), we see that the system is an insulator, which is
confirmed by compressibility data.

In Lieb's original proof, the on-site repulsion was assumed to be uniform in order to satisfy particle-hole symmetry. 
However, with the manifestly symmetric form of the Hubbard Hamiltonian considered here, this restriction is removed, and the system is particle-hole symmetric at half filling for any distribution of $U_{\mathbf{i}}$ through the lattice sites $\mathbf{i}$. 
Further, the strong coupling limit needed to extend the proof to the inhomogeneous lattice is provided by a subsequent work,\cite{Franca10} which established that when two adjacent sites had different values of $U$, say $U_p\neq U_d$, the exchange coupling becomes $4t^2/\widetilde{U}$, with $\widetilde{U}$ being the geometric mean between $U_p$ and $U_d$, provided single occupancy could be enforced in this limit.
Therefore, ferromagnetism is also expected to occur when $U_p\neq U_d >0$.

However, this strong coupling limit breaks down when either $U_p$ or
$U_d$ vanishes so we have also examined this situation.  From our QMC
simulations we established that switching off $U_d$ preserves the
ferromagnetic state with the same main features of the homogeneous case,
while switching off $U_p$ suppresses ferromagnetism (or any other
magnetically ordered state).  Once again, the different environments of
the sites with non-zero repulsion is responsible for this: when $U_p=0$
the system is metallic, and single occupancy of the $p$ sites is no
longer guaranteed.

We have also considered doping away from half filling.   An interesting
feature develops in the $dd$ spin correlations
when $U_p=0$ strong antiferromagnetic correlations;  they attain
a large negative value at $\rho=1/3$, 
caused by occupancy of each $d$ site by a single fermion.
Previous studies\cite{Scalettar91} examined the occupations, local 
moments and pairing for a range of $\epsilon_p - \epsilon_d$,
including $\epsilon_p - \epsilon_d=0$,
but the sharp feature in the dd spin correlations 
in this case was not noted.

In closing, we should mention that the quantitative exploration of
itinerant ferromagnetism remains a key area of strongly correlated
electron systems.  Recently, ferromagnetism has also been observed in
the absence of a lattice in mixtures of $^6$Li atoms in two hyperfine
states.\cite{jo09} Lattice models remain more challenging for such
optical lattice emulation, owing to the difficulty in cooling the atoms
below the ordering temperature, and because of the density inhomogeneity
introduced by the confining potential.  Progress in observing
antiferromagnetism in the single band Hubbard model in one,\cite {Boll16} two,\cite {Parsons16, Cheuk16} and three\cite{hart15} dimensions is ongoing. Because of the tunability of these cold atom
systems, and particularly the fact that different geometries and regimes
of very large $U$ can be accessed, it is possible that new insight into
Hubbard model ferromagnetism is on the horizon.

%%%%%%%%%%%%%%%%%%%%%%%%%%%%%%%%%%%%%%%%%%%%%%%%%%%%%%%%%%%%%%%%%%
%%%%%%%%%%%%%%%%%%%%%%%%%%%%%%%%%%%%%%%%%%%%%%%%%%%%%%%%%%%%%%%%%%
\section*{ACKNOWLEDGMENTS}
%%%%%%%%%%%%%%%%%%%%%%%%%%%%%%%%%%%%%%%%%%%%%%%%%%%%%%%%%%%%%%%%%%
%%%%%%%%%%%%%%%%%%%%%%%%%%%%%%%%%%%%%%%%%%%%%%%%%%%%%%%%%%%%%%%%%%

The work of RTS was supported by the Department of Energy, DOE grant
number DE-SC0014671.  Financial support from the Brazilian Agencies
CAPES, CNPq,  FAPERJ and Science Without Borders Program is also gratefully acknowledged.

%%%%%%%%%%%%%%%%%%%%%%%%%%%%%%%%%%%%%%%%%%%%%%%%%%%%%%%%%%%%%%%%%%%%%%%%
%%%
%%%%%     BIBLIOGRAPHY
%%%%%%%%%%%%%%%%%%%%%%%%%%%%%%%%%%%%%%%%%%%%%%%%%%%%%%%%%%%%%%%%%%%%%%%%%%%

%\begin{thebibliography}{100}

\bibliography{bib_Lieb}
\end{document}